\documentstyle[12pt,epsf]{article}
\chardef\letterchar=11
\chardef\otherchar=12
\chardef\eolinechar=5
\ifx\selectfont\undefined%
  \def\mathrm#1{{\rm #1}}
\fi
\def\ifmath#1{\relax\ifmmode #1\else $#1$\fi}%
\let\ifmathx0
\def\fixmath{\def\ifmath{\noexpand\ifmathx}}%

\newcount\hrs\newcount\minu\newcount\temptime
\def\hm{\hrs=\time \divide\hrs by 60 \minu=\time\temptime=\hrs
\multiply\temptime by 60%
\advance\minu by -\temptime
\ifnum\minu<10 \let\zerofill=0\else \let\zerofill=\relax\fi
 \the\hrs:\zerofill\the\minu}%
\def\ra{\ifmath{\rightarrow}}

\def\lapprox{\ifmath{\sim\kern-1em\raise 0.65ex\hbox{$<$}}}
\def\rapprox{\ifmath{\sim\kern-1em\raise 0.65ex\hbox{$>$}}}
\def\gam{\ifmath{\gamma}}%
\def\MZ{\ifmath{m_{\mathrm{Z}}}}%
\def\antibar#1{\ifmath{#1\bar{#1}}}%

\def\bbbar{\antibar{\mathrm{b}}}%

\def\ccbar{\antibar{\mathrm{c}}}%

\def\ssbar{\antibar{\mathrm{s}}}%

\def\uubar{\antibar{\mathrm{u}}}%

\def\ddbar{\antibar{\mathrm{d}}}%

\def\qqbar{\antibar{\mathrm{q}}}%

\def\nnbar{\antibar{\nu}}%
\def\ee{\ifmath{\mathrm{e^+ e^-}}}%
\def\mm{\ifmath{\mathrm{\mu^+ \mu^-}}}%
\def\tautau{\ifmath{\mathrm{\tau^+ \tau^-}}}%

\def\Zo{\ifmath{\mathrm {Z}}}

\def\GZ{\ifmath{\Gamma_{\mathrm{Z}}}}

\def\oalfs{\ifmath{{\cal O}(\alpha^2) \:}}
\def\ee{\ifmath{\mathrm{e^+ e^-}}}%
\def\mm{\ifmath{\mathrm{\mu^+ \mu^-}}}%
\def\tautau{\ifmath{\mathrm{\tau^+ \tau^-}}}%

\def\Zo{\ifmath{\mathrm {Z}}}

\def\GZ{\ifmath{\Gamma_{\mathrm{Z}}}}

\def\oalfs{\ifmath{{\cal O}(\alpha^2) \:}}
\def\xb{\ifmath{\raise.4ex\hbox{$\chi$}_{_{\mathrm{B}}}}}
\def\x{\raise.4ex\hbox{$\chi$}}

\def\chino{\ifmath{\mathchoice%
     {\displaystyle\raise.4ex\hbox{$\displaystyle\tilde\chi$}}%
        {\textstyle\raise.4ex\hbox{$\textstyle\tilde\chi$}}%
      {\scriptstyle\raise.3ex\hbox{$\scriptstyle\tilde\chi$}}%
{\scriptscriptstyle\raise.3ex\hbox{$\scriptscriptstyle\tilde\chi$}}}}

\let\etaa=\eta
\def\eta{\ifmath{\etaa}}%
\let\iotaa=\iota%
\def\iota{\ifmath{\iotaa}}%
\def\TeV{\ifmmode \hbox{\rm\ Te\kern -0.1em V}\else
                  \hbox{\mathrm{Te\kern -0.1em V}}\fi}%
\def\GeV{\ifmmode \hbox{\rm\ Ge\kern -0.1em V}\else
                  \hbox{\mathrm{Ge\kern -0.1em V}}\fi}%
\def\MeV{\ifmmode \hbox{\rm\ Me\kern -0.1em V}\else
                  \hbox{\mathrm{Me\kern -0.1em V}}\fi}%
\def\keV{\ifmmode \hbox{\rm\ ke\kern -0.1em V}\else
                  \hbox{\mathrm{ke\kern -0.1em V}}\fi}%
\def\eV{\ifmmode \hbox{\rm\ e\kern -0.1em V}\else
                 \hbox{\mathrm{e\kern -0.1em V}}\fi}%
\def\mrad{\ifmmode \hbox{\rm\ mrad}\else
                 \hbox{\mathrm{mrad}}\fi}%

\def\mum{\ifmmode \hbox{\rm $\mu$\kern -0.1em m}\else
                  \hbox{\mathrm{$\mu$\kern -0.1em m}}\fi}%
\def\MM{\ifmmode \hbox{\rm m \kern -0.2em m}\else
                  \hbox{\mathrm{m\kern -0.2em m}}\fi}%
\newbox\boxsqbox
\newdimen\boxsize\boxsize=1.2ex%
\def\boxop{%
\setbox\boxsqbox=\vbox{\hrule depth0.8pt width0.8\boxsize height0pt%
                       \kern0.8\boxsize
                       \hrule height0.8pt width0.8\boxsize depth0pt}%
           \hbox{%
           \vrule height1.0\boxsize width0.8pt depth0pt%
           \copy\boxsqbox
           \vrule height1.0\boxsize width0.8pt depth0pt\kern1.5pt}}%
\def\pmb#1{\setbox0=\hbox{$#1$}
  \kern-.025em\copy0\kern-1.0\wd0%
  \kern.05em\copy0\kern-1.0\wd0%
  \kern-.025em\raise.0433em\box0}%
\newdimen\dkwidth
\def\dk{%
   \dkwidth=\baselineskip
   {\def\to{\rightarrow}
   \kern 3pt%
   \hbox{%
      \raise 3pt%
      \hbox{%
         \vrule height 0.8\dkwidth width 0.7pt depth0pt%
      }%
      \kern-0.4pt%
      \hbox to 1.5\dkwidth{%
         \rightarrowfill
      }%
   \kern0.6em%
   }}%
}%
\def\eqalign#1{%
   \,
   \vcenter{%
      \openup\jot\m@th
      \ialign{%
         \strut\hfil$\displaystyle{##}$&&$%
         \displaystyle{{}##}$\hfil\crcr#1\crcr%
      }%
   }%
   \,
}%
\newcommand {\zfi}{$_{Z\!F}\!I\!^{\textstyle T}\!\!T\!\!_{{\textstyle E}\!R}$}
\newcommand{\sma}{SMATASY}

\newcommand{\alr}{A_{lr}}
\def\egvf{\ifmath{\hat{g}_{\mathrm{V}}^f}}
\def\egaf{\ifmath{\hat{g}_{\mathrm{A}}^f}}

\def\egve{\ifmath{\hat{g}_{\mathrm{V}}^{\mathrm{e}}}}
\def\egae{\ifmath{\hat{g}_{\mathrm{A}}^{\mathrm{e}}}}

\def\sgvf{\ifmath{(\hat{g}_{\mathrm{V}}^f)^2}}
\def\sgaf{\ifmath{(\hat{g}_{\mathrm{A}}^f)^2}}

\def\sgve{\ifmath{(\hat{g}_{\mathrm{V}}^{\mathrm{e}})^2}}
\def\sgae{\ifmath{(\hat{g}_{\mathrm{A}}^{\mathrm{e}})^2}}

\newcommand{\imag}{\mbox{$\Im$m}}
\newcommand{\real}{\mbox{$\Re$e}}

\def\afb{\ifmath{\mathrm {{\cal A}_{fb}}}}

\def\apol{\ifmath{\mathrm {{\cal A}_{pol}}}}
\def\afbpol{\ifmath{\mathrm {{\cal A}_{fb-pol}}}}

\def\gmu{\ifmath{\mathrm{G}_\mu}}
\def\ovgmu{\ifmath{\overline{\mathrm{G}}_\mu}}

\def\xstot{\ifmath{\sigma_{\mathrm{tot}}}}

\def\xspol{\ifmath{\sigma_{\mathrm{pol}}}}

\def\ta1{\ifmath{\tau \rightarrow {\mathrm a_1} \nu}}

\newcommand{\bc}{\begin{center}}
\newcommand{\ec}{\end{center}}
\newcommand{\bq}{\begin{equation}}
\newcommand{\eq}{\end{equation}}
\newcommand{\ba}{\begin{eqnarray}}
\newcommand{\ea}{\end{eqnarray}}
\newcommand{\ban}{\begin{eqnarray*}}
\newcommand{\ean}{\end{eqnarray*}}

\newcommand {\zf}{ZFITTER\ }

\def\ovMZ{\ifmath{{\overline m}_Z}}
\def\ovGZ{\ifmath{{\overline \Gamma}_Z}}

\def\jtf{\ifmath{j^f_{\mathrm{tot}}}}
\def\rtf{\ifmath{r^f_{\mathrm{tot}}}}
\def\jaf{\ifmath{j^f_{\mathrm{A}}}}
\def\raf{\ifmath{r^f_{\mathrm{A}}}}

\def\rfbf{\ifmath{r^f_{\mathrm{fb}}}}
\def\jfbf{\ifmath{j^f_{\mathrm{fb}}}}
\def\rpolf{\ifmath{r^f_{\mathrm{pol}}}}
\def\jpolf{\ifmath{j^f_{\mathrm{pol}}}}
\def\rfbpolf{\ifmath{r^f_{\mathrm{fb-pol}}}}
\def\jfbpolf{\ifmath{j^f_{\mathrm{fb-pol}}}}

\def\Rgf{\ifmath{\mathrm {R}^f_\gam}}

\def\RZfi{\ifmath{\mathrm {R_Z}^{fi}}}

\def\Ffin{\ifmath{\mathrm {F}_n^{fi}}}

\def\RZfn{\ifmath{\mathrm {R_Z}^{f0}}}
\def\RZfe{\ifmath{\mathrm {R_Z}^{f1}}}
\def\RZfz{\ifmath{\mathrm {R_Z}^{f2}}}
\def\RZfd{\ifmath{\mathrm {R_Z}^{f3}}}

\newcommand{\lone}{LEP~1}

\newcommand{\nll}{\nonumber \\}

\def\afb{\ifmath{{\cal A}_{FB}}}
\def\apol{\ifmath{{\cal A}_{pol}}}
\def\afbpol{\ifmath{{\cal A}_{\mbox{\scriptsize \it FB-pol}}}}
\def\alr{\ifmath{{\cal A}_{lr}}}
\def\xslr{\ifmath{\sigma_{lr}}}
\def\afblr{\ifmath{{\cal A}_{\mbox{\scriptsize \it FB-lr}}}}
\def\alrpol{\ifmath{{\cal A}_{\mbox{\scriptsize \it lr-pol}}}}
\def\xslrpol{\ifmath{\sigma_{\mbox{\scriptsize \it lr-pol}}}}
\def\xstot{\ifmath{\sigma_T}}
\def\ovMZ{\ifmath{{\overline M}_Z}}
\def\MZ{\ifmath{M_Z}}
\def\rtf{\ifmath{r^f_{T}}}
\def\jtf{\ifmath{j^f_{T}}}
\def\rfbf{\ifmath{r^f_{FB}}}
\def\jfbf{\ifmath{j^f_{FB}}}
\def\rpolf{\ifmath{r^f_{pol}}}
\def\jpolf{\ifmath{j^f_{pol}}}
\def\rfbpolf{\ifmath{r^f_{\mbox{\scriptsize \it FB-pol}}}}
\def\jfbpolf{\ifmath{j^f_{\mbox{\scriptsize \it FB-pol}}}}
\def\rfblrf{\ifmath{r^f_{\mbox{\scriptsize \it FB-lr}}}}
\def\jfblrf{\ifmath{j^f_{\mbox{\scriptsize \it FB-lr}}}}
\def\rlrpolf{\ifmath{r^f_{\mbox{\scriptsize \it lr-pol}}}}
\def\jlrpolf{\ifmath{j^f_{\mbox{\scriptsize \it lr-pol}}}}
\def\rlrf{\ifmath{r^f_{lr}}}
\def\jlrf{\ifmath{j^f_{lr}}}
\def\ttbar{\antibar{\mathrm{t}}}%
 
\begin{document}
\thispagestyle{empty}
\onecolumn
\begin{flushleft}
{\tt DESY 94-125
\\   July 1994  }
\end{flushleft}
\vspace*{2.50cm}

\LARGE
{
\noindent
SMATASY  
 
 
\noindent
A program for the model independent  \\ description
of the $Z$ resonance
}
 
\vspace{2.5cm}
\noindent
\Large
Stefan~Kirsch\footnote{
Present address: Particle Physics Division, CERN.}
\, and \,
Tord~Riemann
\vspace{.7cm}
\\
\noindent
\normalsize
{\Large \em
DESY -- Institut f\"ur Hochenergiephysik
\vspace{1mm}
\\
Platanenallee 6, D-15738 Zeuthen, Germany
\vspace{1mm}
}

\vspace{\fill}
\vfill
\thispagestyle{empty}
\noindent
{\large
{ABSTRACT 
\vspace*{.2cm}
\\}
}
\noindent
\sma\ is an interface for the \zfi\ package and may be used for 
the model independent description of the $Z$ resonance at LEP~1 and SLC.
It allows the determination of the \Zo\ mass and width and	
its resonance shape parameters $r$ and $j$ for cross-sections and their
asymmetries.
The $r$ describes the peak height and $j$ the interference of the $Z$ 
resonance with photon exchange in each scattering channel and for
$\sigma_T$, $\sigma_{FB}$, $\sigma_{lr}$, $\sigma_{pol}$ etc. separately.
Alternatively, the helicity amplitudes for a given scattering channel may be
determined. 
We compare our formalism with other model independent approaches.
The model independent treatment of QED corrections in \sma\ is applicable also
far away from the \Zo\ peak.
\normalsize
\vspace*{.5cm}
\newpage
{\Large \bf PROGRAM SUMMARY}

\begin{description}
\item[\it Title of program:] SMATASY version 2.2 

\item[\it Catalogue number:] 

\item[\it Program obtainable from:] CPC Program Library, Queen's University
of Belfast, N. Ireland (see application form in this issue) 

\item[\it Licensing provisions:] none 

\item[\it Computer for which the program is designed:] any computer with a
FORTRAN 77 compiler 

\item[\it Operating system under which the program has been tested:] UNIX

\item[\it Programming language used:] FORTRAN 77

\item[\it Memory required to execute with typical data:] 34000 words

\item[\it No. of bits in a word:] 32

\item[\it No. of lines in distributed program, including test data, etc:]
1204

\item[\it Keywords:] electron--positron annihilation, scattering matrix,
cross sections, asymmetries, model independent ansatz,
radiative corrections

\item[\it Nature of the physical problem:] 
The program determines cross sections and asymmetries in the vicinity of the
\Zo\ resonance taking into account radiative corrections up to \oalfs.
This model independent determination is based on the S-matrix
approach.

\item[\it Methode of solution:] 
SMATASY is designed as a new interface of the \zf\ package \cite{zfitter}.
Therefore it relies on the same treatment of radiative corrections as \zf.

\end{description}
\vspace{0.5cm}

{\Large \bf LONG WRITEUP}

\section {
Introduction
\label {sec1}
}
With the rising precision of the experimental study
of the $Z$ resonance at LEP 1 and SLC, an
interpretation of the data in a rigorously model independent way becomes more
and more important.
A wrong theoretical description of the data may lead to systematic
non-observed
shifts of the measured parameters which describe
the $Z$.
 
\sma\ is a program for the model independent study of
fermion pair production
at the \Zo\ resonance:
\bq
e^+ e^- \rightarrow (\gamma, Z) \rightarrow f^+ f^- (n \gamma).
\label{e0}
\eq
Under the assumption that
QED and QCD are well understood theories,
the
total cross-section around the \Zo\ peak may be characterized by four
                                                            real
parameters: the mass
$M_Z$, width $\Gamma_Z$, the residuum $r$ of the $Z$ resonance, and the 
strength of the \gam\Zo-interference $j$.
The parameters $r$ and $j$ are related to helicity amplitudes \RZfi\
for $e^+ e^-$ annihilation into fermion pairs via \Zo-exchange.
Further,
the measured cross-section includes photonic virtual and real
bremsstrahlung corrections which may be described by a flux function
$\rho_T(s'/s)$~\cite{smatrix}:
\ba
\sigma_T(s) = 
\frac{4}{3} \pi \alpha^2
\int \frac{ds'}{s} 
\left[ \frac{r^{\gamma}}{s} +
\frac {s r + (s - M_Z^2) j} {(s-M_Z^2)^2 + M_Z^2 \Gamma_Z^2}
\right]
\rho_T
\left(\frac{s'}{s}\right).
\label{sigqed}
\ea
As mentioned, the photon exchange parameter
$r^{\gamma}$ is assumed to be known, and $s$ depends on the beam energy,
$s=4 E^2$.
Further, at \lone\ and SLC very precise
measurements of various cross-section asymmetries are performed.
An example is shown in figure~\ref{asy}.
In the vicinity of the \Zo\ peak,
these asymmetries behave relatively smoothly and
may be described by a simple, universal   formula~\cite{sma2}:
\ba
A(s) = A_0 + C(s) \,  A_1 \left( \frac{s}{M_Z^2}-1 \right).
\label{asyqed}
\ea
The QED corrections are contained in the factor $C(s)$.

\bigskip
 
The model independent description of cross-sections around the \Zo\ resonance
with account of QED corrections may be done with the 
\zfi~\cite{zfitter}--\cite{plb229} branch {\tt ZUSMAT}.
Here, we describe the Fortran program \sma\ which is
designed as an interface to \zfi\
for the model independent description of asymmetries.
\sma\ provides
the full
functionality of \zfi\ with all its possibilities of flag settings,
different treatments of photonic bremsstrahlung and
 QCD
corrections.
 
\sma\ is devoted to the following tasks:
\begin{itemize}
\item
determination of the \Zo-exchange parameters $r$ and
\gam\Zo-interference parameters $j$ for
cross-sections and asymmetries,
\item
determination of the asymmetry parameters $A_0$ and $A_1$,
\item
determination of the couplings \RZfi\ of helicity amplitudes describing the 
\Zo-exchange matrix element,
\item
model independent treatment
of QED corrections for cross-sections and asymmetries.
\end{itemize}

The basic formulae are introduced in section~2 and related to other approaches
in section~3.
The structure of the \sma\ package is explained in section~4 while the
procedures are described in section~5.
Appendix~A contains a sample output of \sma. 

\def\swid{0.8\textwidth}
\begin{figure}[thbp]
\begin{center}
\mbox{\epsfxsize=\swid\epsffile {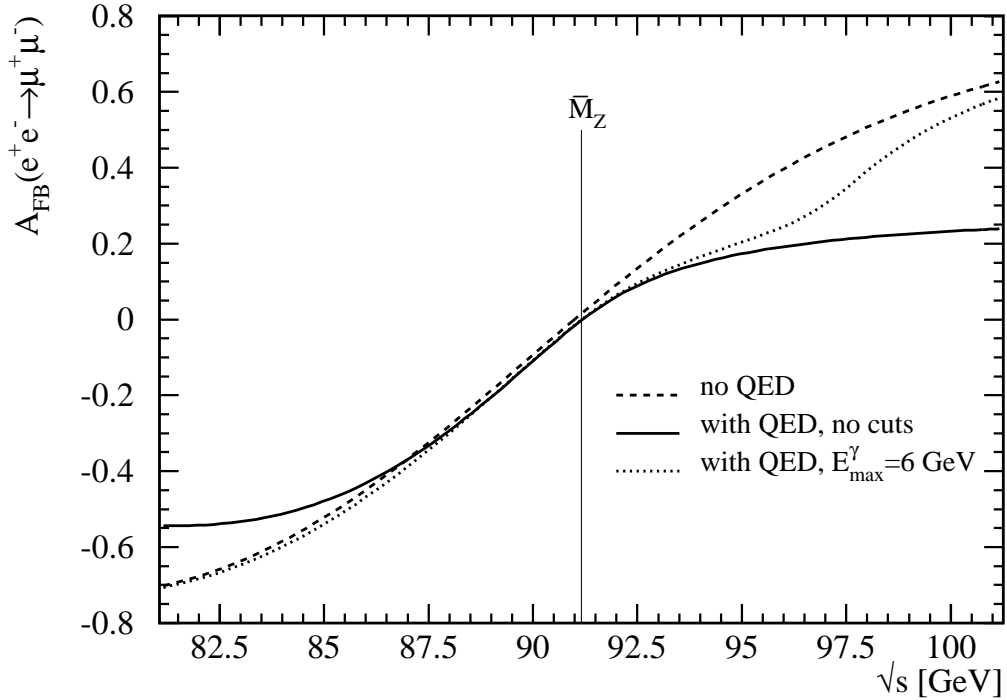}}
\end{center}
\caption[foo]{
The forward-backward asymmetry for the process $e^+ e^- \rightarrow \mu^+
\mu^-$ near the \Zo\ peak.
\label{asy}
}
\end{figure}

\section {
Basic formulae
\label{sec:smatrix}
}
\subsection {
Cross-sections at the \Zo\ resonance
\label{cs}
}
The matrix element for the production of a fermion
pair near the \Zo\ resonance may be described by four helicity amplitudes.
In full generality, they have the following form~\cite{smatrix,stuart}:
\ba
\label{eqn:mat0}
{\cal M}^{fi}(s) = \frac{\Rgf}{s} + \frac{\RZfi}{s-s_Z} +
\sum_{n=0}^\infty \frac{\Ffin}{\ovMZ^2} \left(\frac{s-s_Z}{\ovMZ}\right)^n 
,~~\  i=1,\ldots,4.
\ea
The position of the \Zo\ pole in the complex $s$ plane is given by 
$s_{\Zo}$:
\ba
\label{szb}
s_Z = \ovMZ^2 - i \ovMZ \ovGZ.
\ea
The \Rgf\ and \RZfi\ are complex residua of the photon and the \Zo\ boson,
respectively.
In Born approximation, they are real numbers.
In the Standard Model, contributions from higher order
corrections are incorporated and the residua become complex numbers.
The coefficients \Ffin\
in~(\ref{eqn:mat0}) describe
nonresonant contributions to the scattering process.
In the Standard Model they arise from higher order corrections, e.g. from
\Zo\Zo\ and WW box diagrams.
The s dependence of the virtual electroweak corrections also contributes 
to them.
 
To a very good approximation, the Taylor series in~(\ref{eqn:mat0})
may be neglected.
Its first term F$_0^{fi}$ is constant.
In a \Zo\ line shape fit, it will be strongly correlated
with the photon exchange term $
\Rgf /s \approx \Rgf /\ovMZ^2$,
\ba
\Rgf = \displaystyle{ F_A(s) \left| Q_e Q_f \right|}.
\label{gamma}
\ea
The
$Q_f$ is the electric charge of the final state fermions and
$F_A(s)$ the vacuum polarization of the photon:
\ba
F_A(\ovMZ^2) = \displaystyle{\frac{\alpha(\ovMZ^2)}{\alpha}}
=
\frac{137.06}{128.86} - i \, 0.0188.
\ea
In practice it seems to be impossible to disentangle $F_A$ and the
nonresonating quantum correction F$_0^{fi}$.
The next-to-leading background
 term is proportional to
$\sigma = (s-s_Z)/\ovMZ^2 
\approx 2 (\sqrt{s}/\ovMZ-1)$.
At LEP~1 and SLC the bulk of data is taken at
$|\sqrt{s}-\ovMZ| < \ovGZ$ and $\sigma$ becomes less than 
5 \%.
Having in mind that the $F_n^{fi}$ are quantum corrections and proportional to
$\alpha / \pi$, it is for all practical needs:
\ba
\label{eqn:mat}
{\cal M}^{fi}(s) 
&=& \frac{\Rgf}{s} + \frac{\RZfi}{s-s_Z} +
\frac{F_0^{fi}}{\ovMZ}
\nll
&\approx&~
\frac{\Rgf}{s} + \frac{\RZfi}{s-s_Z} .
\ea
There are four residua \RZfi\ for the
four independent helicity amplitudes in the case of massless external
fermions:
\ba
\renewcommand{\arraystretch}{1.2}
\label{eqn:mi_he14}
\begin{array}{lcl}
R_Z^{f0} & = & R_Z (e^-_Le^+_R \longrightarrow f^-_L f^+_R), \\
R_Z^{f1} & = & R_Z (e^-_Le^+_R \longrightarrow f^-_R f^+_L), \\
R_Z^{f2} & = & R_Z (e^-_Re^+_L \longrightarrow f^-_R f^+_L), \\
R_Z^{f3} & = & R_Z (e^-_Re^+_L \longrightarrow f^-_L f^+_R).
\end{array}
\renewcommand{\arraystretch}{1.}
\ea
The amplitudes
${\cal M}^{fi}(s)$ give rise to four cross-sections $\sigma_i$
which add up incoherently to the following measurable cross-sections:
\ba
\renewcommand{\arraystretch}{1.2}
\label{eqn:xs14}
\begin{array}{lclcl}
&&\sigma_{T}^0(s)      & = & +~  \sigma_0 + \sigma_1 + \sigma_2 + \sigma_3,
  \\
\sigma_{\mbox{\scriptsize \it lr-pol}}^0(s) & = &
\sigma_{FB}^0(s)     & = & +~  \sigma_0 - \sigma_1 + \sigma_2 - \sigma_3,
 \\
\sigma_{\mbox{\scriptsize \it FB-lr}}^0(s)& = &
\sigma_{pol}^0(s)    & = & -~  \sigma_0 + \sigma_1 + \sigma_2 - \sigma_3
, \\
\sigma_{lr}^0(s)& = &
\sigma_{\mbox{\scriptsize \it FB-pol}}^0(s) & = & -~  \sigma_0 -
\sigma_1 + \sigma_2 + \sigma_3.
\end{array}
\renewcommand{\arraystretch}{1.}
\ea
Here, the 
$\sigma_{T}^0$ is the total cross-section, $\sigma_{FB}^0$ defines the
forward-backward asymmetry, $\sigma_{pol}^0$ the final state
polarization,
$\sigma_{FB,pol}^0$ the forward-backward asymmetry of the final state
polarization 
etc.\footnote{The agreements between several pairs of asymmetries will be
disturbed after inclusion of QED corrections, see
below.} 
 
All these cross-sections may be parameterized by the following master
formula~\cite{smatrix}: 
\ba
\renewcommand{\arraystretch}{2.2}
\label{eqn:smxs}
\nonumber
\sigma_A^0(s)
&=&
\displaystyle{\frac{4}{3} \pi \alpha^2
\left[ \frac{r^{\gamma f}_A}{s} +
\frac {s r^f_A + (s - \ovMZ^2) j^f_A} {(s-\ovMZ^2)^2 + \ovMZ^2 \ovGZ^2} 
+ \frac{r_A^{f0}}{\ovMZ^2}\right]},
\hspace{.5cm} A = \mbox{\it T, FB, pol, FB-pol,} \ldots \\
&\approx&
\displaystyle{\frac{4}{3} \pi \alpha^2
\left[ \frac{r^{\gamma f}_A}{s} +
\frac {s r^f_A + (s - \ovMZ^2) j^f_A} {(s-\ovMZ^2)^2 + \ovMZ^2 \ovGZ^2}
\right]}
.
\renewcommand{\arraystretch}{1.}
\ea
The $r^{\gamma f}_A$ is the photon exchange term,
\ba
\label{qedr}
r^{\gamma f}_A =
\displaystyle{\frac{1}{4}  c_f \sum_{i=0}^3 \{\pm 1\}
\left| \Rgf \right|^2 } R_{QCD},
\ea
vanishes for all asymmetric cross-sections.
$c_f=1,3$ for leptons and quarks, respectively.
QCD corrections for quarks are taken into account by the factor $R_{QCD}$ of
\cite{zfitter}.
The \Zo-exchange residuum $r^f_A$ and the \gam\Zo-interference
$j^f_A$ are:
\ba
\label{eqn:rrjj}
\renewcommand{\arraystretch}{2.}
\begin{array}{lll}
r^f_A & = & \displaystyle{c_f \left\{\frac{1}{4}\sum_{i=0}^3 \{\pm 1\}
\left| \RZfi \right|^2 + 2 \frac{\ovGZ}{\ovMZ} \imag C^f_A \right\}} R_{QCD},
\\ 
j^f_A & = & c_f\left\{2 \real C^f_A - 2 \displaystyle{\frac{\ovGZ}{\ovMZ}}
\imag C^f_A\right\} R_{QCD},
\\
C^f_A & = & (R^f_\gamma)^* \left(\displaystyle{\frac{1}{4}\sum_{i=0}^3
\{\pm 1\} \RZfi}\right) .
\end{array}
\renewcommand{\arraystretch}{1.}
\ea
The factors $\{\pm 1\}$ in~(\ref{qedr}) and~(\ref{eqn:rrjj})
indicate that the signs of $\left| R_{\gamma}^f \right|$, 
$\left| \RZfi \right| ^2$,
and of \RZfi\ correspond to the signs of $\sigma_i$ in~(\ref{eqn:xs14}).
\subsection {
QED corrections for cross-sections
\label{qct}
}
For the calculation of QED corrections in
\sma\ the \zfi\ environment is used.
This is done by convoluting~(\ref{eqn:smxs}) with radiator functions for
initial and final state radiation and their interference.
The initial and final state corrections with soft photon exponentiation
to the cross-sections \xstot, \xspol, \xslr, \xslrpol\, 
are described by:
\ba
\sigma_A(s) = \int \frac{ds'}{s} \sigma_A^0(s') \rho_T\left(\frac{s'}{s}
\right) \hspace{.5cm} A= \mbox{\it T, pol, lr, lr-pol}.
\label{sigqed2}
\ea
Analogously, for the forward-backward difference cross-sections,
\ba
\sigma_a(s) = \int \frac{ds'}{s} \sigma_a^0(s')
\rho_{FB}\left(\frac{s'}{s}\right),
\hspace{.5cm} a = \mbox{\it FB, FB-pol, FB-lr.} 
\label{sigqed3}
\ea
We stress that $\rho_{FB}(s'/s) \neq \rho_T(s'/s)$.
The contributions from initial final state interference bremsstrahlung are
slightly more complex~\cite{zfitter}--\cite{plb229}.
At the \Zo\ resonance they are
numerically tiny as long as no strong cuts are applied.
They will be neglected in the following.
With QED corrections, the master formula may be rewritten as follows
~\cite{sma2}:
\ba
{\bar{\sigma}}_A (s)
 &=& \frac{4}{3} \pi \alpha^2
\left[ \frac{ {\bar r}^{\gamma f}_A}{s} +
\frac {s {\bar r}_A^f + (s - \ovMZ^2) {\bar j}_A^f} {(s-\ovMZ^2)^2 + 
\ovMZ^2
\ovGZ^2}
+
\frac{ {\bar r}_A^{0f} }{ M_Z^2 }
\right]
\nll
&\approx&~ 
\frac{4}{3} \pi \alpha^2
\left[ 
\frac{ 
{\bar r}^{\gamma f}_A
}{s} +
\frac{
s {\bar r}_A^f + (s - \ovMZ^2) {\bar j}_A^f
} {
(s-\ovMZ^2)^2 + \ovMZ^2
\ovGZ^2
} 
\right]
.
\label{e20}
\ea
The barred
parameters contain correction
factors with QED corrections:
\ba
\renewcommand{\arraystretch}{1.2}
\begin{array} {lcl}
{\bar{r}}_T^{\gamma f} &=& C^{\gamma}_T(s) \,\, r_T^{\gamma f},
\\
{\bar{r}}_A^f        &=& C^r_A(s) \,\, r_A^f,
\\
{\bar{j}}_A^f        &=& C^j_A(s) \,\, j_A^f,
\\
{\bar r}_A^{0f}      &=& C_A^0(s)  \,\,  r_A^{0f},
\end{array}
\renewcommand{\arraystretch}{1.}
\ea
where
\ba
\label{eqn:cg}
\renewcommand{\arraystretch}{2.4}
\begin{array}{lcl}
C^{\gamma}_T(s)
&=&
\displaystyle{
{\cal I} \left[ \frac{s}{s'} \right]}, 
\\
\label{eqn:car}
C_A^r(s) 
&=& 
\displaystyle{
{\cal I} \left[ \frac{s'}{s} \, \, 
\frac{(s -\ovMZ^2)^2+\ovMZ^2 \ovGZ^2}
                              {(s'-\ovMZ^2)^2+\ovMZ^2 \ovGZ^2} \right]},
\\
\label{eqn:caj}
C_A^j(s)
&=&
\displaystyle{
{\cal I} \left[
      \frac{s'-\ovMZ^2}{s-\ovMZ^2}
\, \,
\frac{(s -\ovMZ^2)^2+\ovMZ^2 \ovGZ^2}
                              {(s'-\ovMZ^2)^2+\ovMZ^2 \ovGZ^2}
\right]} ,
\\
\label{eqn:ca0}
C_A^0(s)
&=&
\displaystyle{
{\cal I} \left[ \frac{(s'-\ovMZ^2)^0}{(s-\ovMZ^2)^0} \right]}
 = {\cal I} \left[ 1 \right] .
\label{e23}
\end{array}
\renewcommand{\arraystretch}{1.}
\ea
Here, the definition 
\bq
{\cal I}_A [B] = \int d\left( \frac{s'}{s} \right) B (s') \rho_A
\left(\frac{s'}{s}\right) 
\label{e22}
\eq
is used.
The QED correction factors are completely model independent, i.e.	
independent of the underlying dynamics of the scattering process.
They {\em depend} on mass and width of the \Zo\ and on
the handling of the
photonic phase space,
the inclusion of higher orders, and on acceptance cuts.
The reader may wonder that the corrections $C_A^{j}$
seem to be singular at
$\sqrt{s}=\ovMZ.$
This is not the case for the {\em products} $C_A^{j}(s) (s-\ovMZ)$ which are
physically relevant.
As may be seen from the corresponding definitions,
these products remain small (but potentially non-vanishing) when
$\sqrt{s}$ approaches \ovMZ.
There the QED corrected cross-sections may be defined as (smooth) limits
from the neighboring energies.

The correction factors are shown in figures~\ref{sm3}--\ref{sm2} without  and
with two different cuts on the photon phase space. 
All QED corrections are smooth and, with one exception, rather independent of
$s$. 
Those to the \Zo\ exchange, $C_A^r$, develop
the radiative tail at the right hand side of the peak which at some value of 
$s$ gets suppressed by the cuts. 
It may be further seen that the corrections to total cross-sections and those
to forward-backward differences are not equal, although 
of similar size near the peak.
They deviate more when more hard bremsstrahlung is possible~\cite{plb229}.
This explains the rise of their difference with the tail and the subsequent  
vanishing of it after the cuts become influential.

\subsection {
Asymmetries around the \Zo\ resonance
\label{a}
}
Without QED corrections, asymmetries are defined by:
\ba
\label{eqn:mi_asy}
{\cal A}_A^0(s) = \frac{\sigma_A^0(s)}{\sigma_{T}^0(s)},~~~ A \neq T.
\ea
These asymmetries take an extremely simple form around the \Zo\
resonance~\cite{sma2}: 
\bq
{\cal A}_A^0(s) = A_0^A + A_1^A \left(\frac{s}{\ovMZ^2} - 1 \right) +
             A_2^A \left(\frac{s}{\ovMZ^2} - 1 \right)^2 + \ldots
\approx
A_0^A + A_1^A \left(\frac{s}{\ovMZ^2} - 1 \right).
\label{e1}
\eq
 
\def\swid{0.8\textwidth}
\begin{figure}[htbp]
\begin{center}
\mbox{\epsfxsize=\swid\epsffile {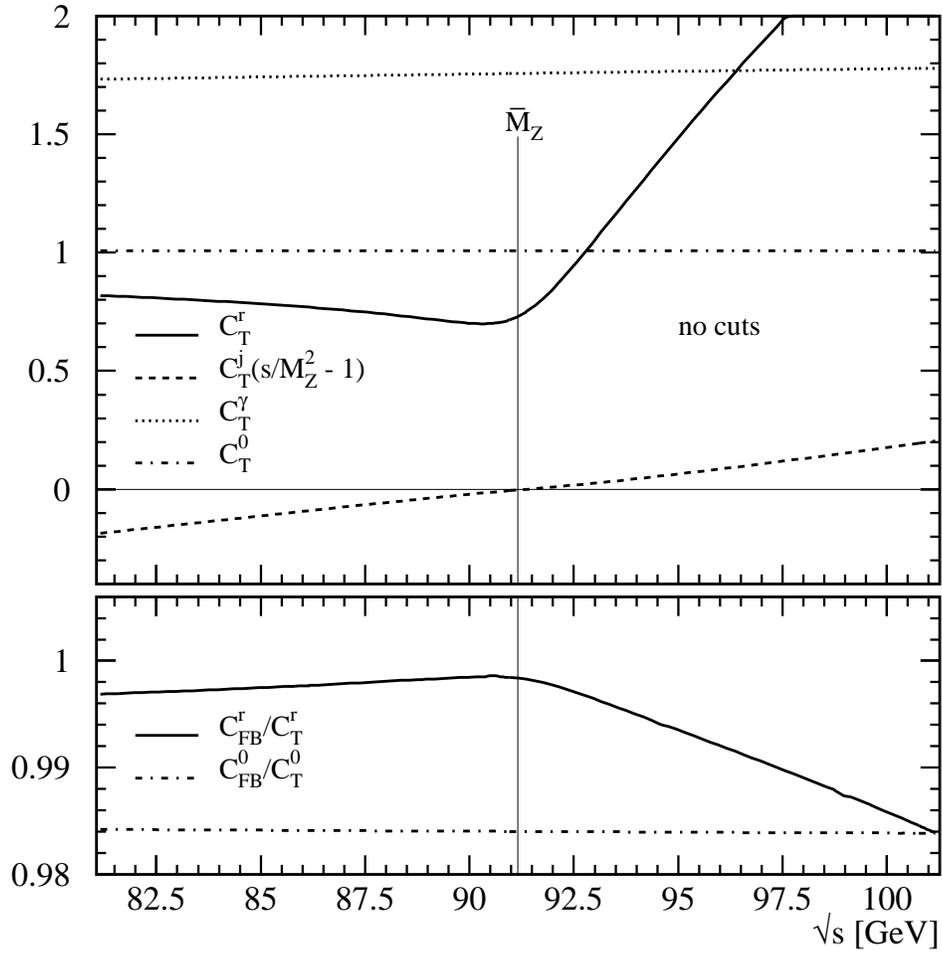}}
\end{center}
\vskip -4.0cm
\caption[foo]{  
The model independent QED correction factors without cuts.
\label{sm3}
}
\end{figure}

\begin{figure}[htbp]
\begin{center}
\def\swid{0.8\textwidth}
\mbox{\epsfxsize=\swid\epsffile {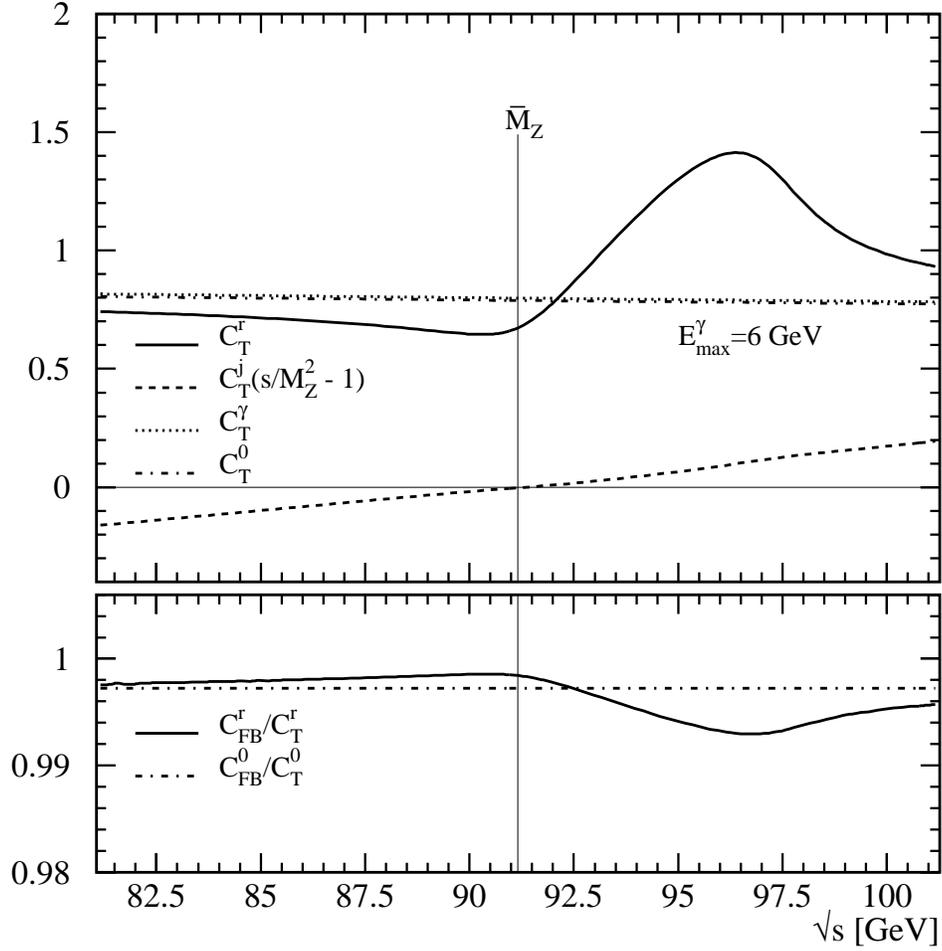}}
\end{center}
\vskip -4.0cm
\caption[foo]{  
The model independent QED correction factors with a cut on the energy of the
bremsstrahlung photon. 
\label{sm7}
}
\end{figure}
 
\begin{figure}[htbp]
\begin{center}
\mbox{\epsfxsize=\swid\epsffile {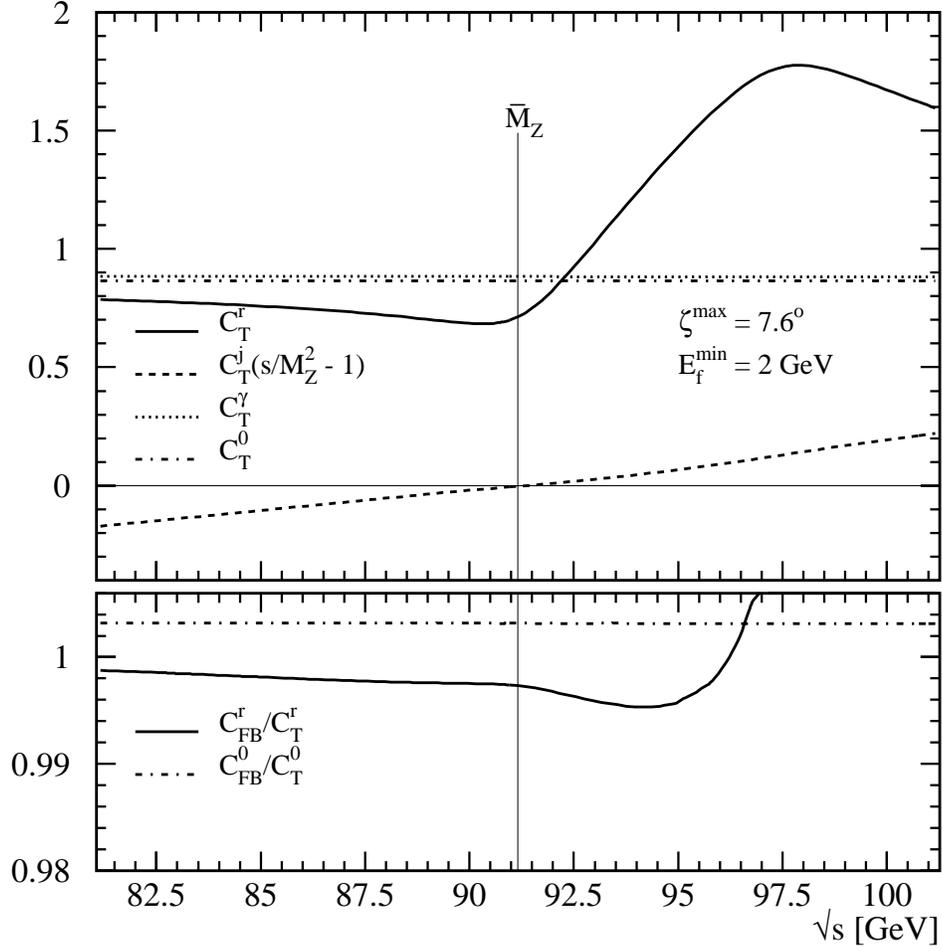}}
\end{center}
\vskip -4.0cm
\caption[foo]{  
The model independent QED correction factors with cuts on the acollinearity 
$\xi$ of
the final state fermions and on their energy $E_{f}$.
\label{sm2}
}
\end{figure}

\noindent
The higher order terms may be safely neglected since $(s/\ovMZ^2-1)^2 =
\sigma^2 < 2 \times 10^{-4}$.
The coefficients have a quite simple form:
\ba
\label{eqn:a0}
A_0^A &=&
\frac{r_A^f}{r_T^f + \gamma^2 r^{\gam f}_T} \approx \frac{r_A^f}
{r_T^f},
\\
\label{eqn:a1}
A_1^A &=&
\left[ \frac{j^f_A}{r_A^f} - \frac{j_T^f - 2\gam^2 r^{\gam f}_T}
{r_T^f+ \gam^2 r^{\gam f}_T} \right] A_0^A
\approx
\left[ \frac{j^f_A}{r_A^f} -  \frac{j_T^f}{r_T^f} \right] A_0^A.
\label{e15}
\ea
Here, the $r_A^{0f}$ is neglected in both $A_0$ and $A_1$. 
Further, the definition
$\gamma^2 = \ovGZ^2 / \ovMZ^2 \approx 0.75 \times 10^{-3}$ is used.
\subsection {
QED corrections for asymmetries
\label{qca}
}
A typical cross-section asymmetry with QED corrections
is shown in figure~\ref{asy}.
The Born asymmetry is linear around the \Zo\ resonance.
QED corrections lead to deviations from this; specially pronounced at
the right hand side of the peak.
Although asymmetries get much smaller QED corrections than the
cross-sections themselves,
an analysis of the data would not be consistent without their correct
treatment.    
 
With QED corrections,
the master formula for asymmetries becomes:
\ba
{\bar{\cal A}}_A(s) 
&=& 
{\bar A}_0^A + {\bar A}_1^A \left(\frac{s}{\ovMZ^2} - 1 \right).
\label{e24}
\ea
The coefficients ${\bar A}_n$
may be obtained from the $A_n$ defined in section~\ref{a} by
replacing in their
definitions the unbarred variables by barred ones.
 
For the peak contributions to the forward-backward asymmetries the
explicit expressions are:
\ba
{\bar A}_0^a
&=& 
\frac{C_{FB}^{r}}{C_T^{r}}
\frac{r_a^f}{r_T^f + [C_T^{\gamma}/ C_T^{r}] 
\gamma^2 r^{\gamma f}_T}
\nll
&\approx&
\frac{C_{FB}^{r}}{C_T^{r}}
\frac{r_a^f}{r_T^f + 0.001}
\nll
&\approx&
0.998 \frac{r_a^f}{r_T^f + 0.001}
\approx \frac{r_a^f}{r_T^f},
     \hspace{.5cm} a = \mbox{\it FB, FB-pol, FB-lr.}
\label{e25}
\ea
Aiming at an experimental accuracy of $\Delta \afb^{lep} = 0.001$ at
LEP~1, the deviation of
$C_{FB}^{r}/C_T^{r}$ from unity has to be taken into account.
Compared to \afb, \afbpol\ and \afblr\ 
the expressions for \apol, \alr\ and \alrpol\ are
simpler since the QED
corrections to the numerator and the denominator are both of the total
cross-section type.
The leading term is:
\ba
{\bar A}_0^{A} &=&
\frac{r_{A}^f}{r_T^f + [C_T^{\gamma}/ C^{r}_{T}] \gamma^2 
r_T^{\gamma f}}
\nll
&\approx&  \frac{r_{A}^f}{r_T^f + 0.001}
\approx  \frac{r_{A}^f}{r_T^f}, \hspace{.5cm} A= \mbox{\it pol, lr, lr-pol}.
\label{e27}
\ea
The coefficients $\bar{A}_1^A$ are in reasonable approximation:
\ba
{\bar{A}_1^a}
&=&
C(s)
\left[ 
 \frac{C_{FB}^{j}}{C_T^{j}} \frac{C_T^{r}}{C_{FB}^{r}}
     \frac{j_a^f}{r_a^f} -  \frac{j_T^f}{r_T^f}
\right] {\bar{A}_0^a}
\nll
&\approx&
C(s)
\left[ 
     \frac{j_a^f}{r_a^f} -  \frac{j_T^f}{r_T^f}
\right] {\bar{A}_0^a}
,
\hspace{.5cm} a=\mbox{\it FB, FB-pol, FB-lr,}
\\
\bar{A}_1^A 
&=&
C(s)
\left[ 
     \frac{j_A^f}{r_A^f} -  \frac{j_T^f}{r_T^f}
\right] {\bar{A}_0^A}, \hspace{.5cm} A=\mbox{\it pol, lr, lr-pol}.
\label{e31a}
\ea
The coefficients with a pronounced $s$ dependence are
\ba
C(s) \equiv C_T(s) = \frac{C^{j}_T}{C^{r}_T},
\hspace{.7cm}
C_{FB}(s) = \frac{C^{j}_{FB}}{C^{r}_{FB}} \approx C(s).
\label{ctfb}
\ea

\begin{figure}[bhtp]
\begin{center}
\def\swid{0.8\textwidth}
\mbox{\epsfxsize=\swid\epsffile {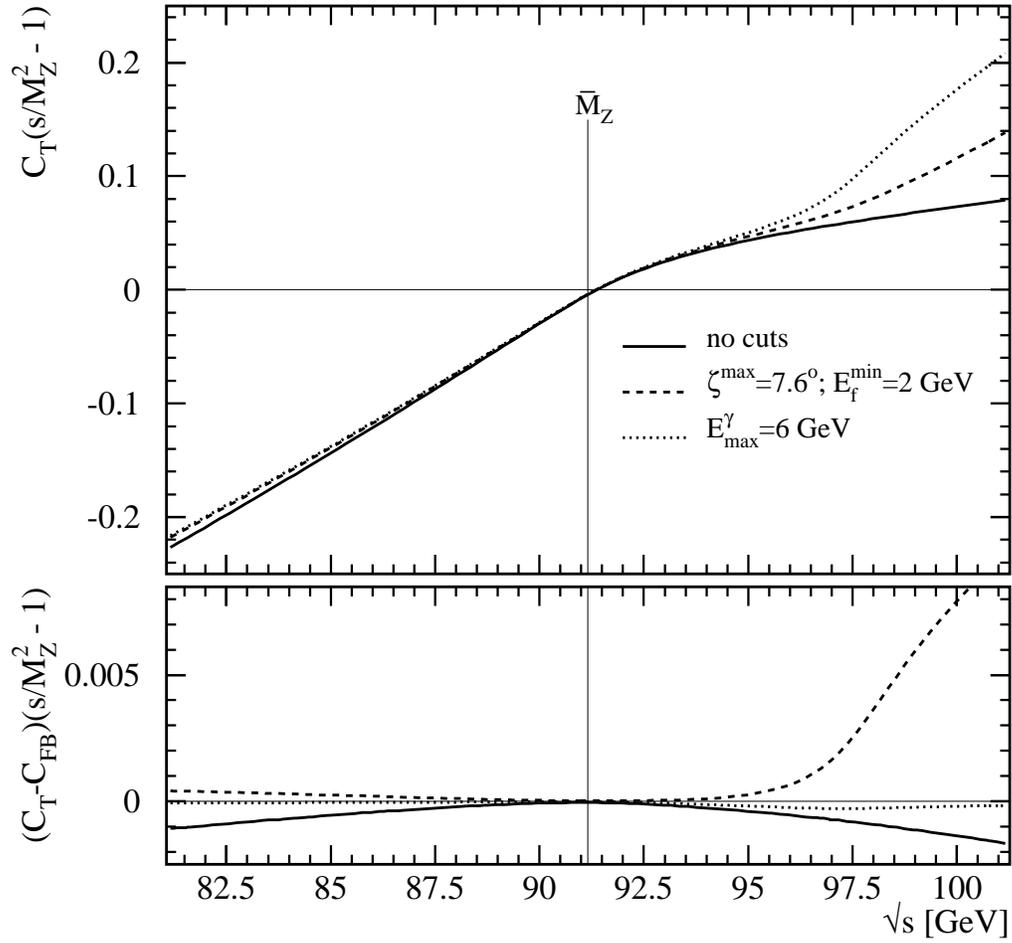}}
\end{center}
\vskip -4.0cm
\caption[foo]{
The model independent QED correction to the slope of asymmetries near the \Zo\
peak. 
\label{sm8}
}
\end{figure}

The behaviour of $C_T$ and $C_{FB}$ 
as functions of $s$ and the dependence on cuts is shown in figure~\ref{sm8}. 
As mentioned already, the $C_T (s/\ovMZ^2-1)$ does not vanish at 
$\sqrt{s}=\ovMZ$ but is extremely small.
Aiming at an accuracy of several per mill, one may neglect the difference
between $C_T$ and $C_{FB}$ in the vicinity of the peak.   
The initial state QED corrections to the \Zo-exchange cross-section
develop a radiative tail while those to the \gam\Zo-interference
do not.
Due to this, their ratio $C(s)$
gets damped at the right hand side of the \Zo\ peak. 
This damping was seen in figure~1. 
It may be assigned to QED corrections completely.
At $\sqrt{s} > \ovMZ$, the radiative tail may be avoided by a cut on the 
energy of the emitted photons:
\bq
\frac{E_{\gamma}}{E_{\mathrm{beam}}} < \Delta = 1 - \frac{M_Z^2}{s}.
\label{e32a}
\eq
At LEP~1 and SLC, where $\sqrt{s} \approx \ovMZ$, this condition is rather
restrictive; e.g. at $\sqrt{s} = \ovMZ + 2 \ovGZ$, it is
$\Delta = 0.1$. 
In figures~1 and~5, photons are cut with energies larger than 6~GeV.
The radiative tail is suppressed by this if $\sqrt{s'} > \ovMZ$ is ensured.
At $\sqrt{s} > \ovMZ / \sqrt{1- E_{\gamma}/E_{\mathrm{beam}} }= 97.5$ GeV this
is the case.
In the immediate vicinity of the peak one unavoidably measures data which
contain radiative corrections.
As may be seen from the figure, the other cuts (on the acollinearity and energy
of the final state fermions) are similar~\cite{zfitter}. 

%

\section{Other model independent approaches}
\subsection{The BCMS approach}
In~\cite{borrelli,jegerl} the following model independent	
formula for the total cross-section has been advocated:
\bq
\displaystyle{
\sigma_T^{o} (s) =
\frac
{12 \pi \Gamma_e \Gamma_f} {\left| s - s_0 \right|^2}
\left\{
\frac{s}{M_Z^2}  +   {\cal R}_f \frac{s-M_Z^2}{M_Z^2}
+ {\cal I}_f \frac{\Gamma_Z}{M_Z} + \ldots
\right\}
+ \sigma_{\mathrm{QED}}^f.
}   
\label{boreli}
\eq
The free parameters of the above expression may be calculated within the
Standard Model with account of electroweak radiative corrections
~\cite{borrelli,isidori}.
There is a simple one--to--one correspondence to the parameters
in~(\ref{eqn:smxs})~\cite{zfitter}:
\ba
\nonumber
r_T^f &=& 
\frac{9}{\alpha^2}  \frac{\Gamma_e\Gamma_f}{M_Z^2}
\left(1+\frac{\Gamma_Z}{M_Z} {\cal I}_f \right), 
\\
j_T^f &=& 
\frac{9}{\alpha^2}  \frac{\Gamma_e\Gamma_f}{M_Z^2}
\left( {\cal R}_f - \frac{\Gamma_Z}{M_Z} {\cal I}_f \right), 
\\
\nonumber
- 2 \left| Q_e Q_f \right| \frac{ \sum \RZfi}{\sum |\RZfi|^2} \Im m F_A
&=& 
{\cal I}_f.
\ea
Further,
\ba
\label{sz0}
s_0 = M_Z^2 - i \frac{s}{M_Z} \Gamma_Z.
\ea
The relation between the definitions of $s_0$ and $s_Z$
is given to a very good approximation
by  the following 
equalities, which also affect the coupling strength~\cite{trafo}:
\ba
\label{mmggzz}
\renewcommand{\arraystretch}{1.2}
\begin{array}{lllllcr}
\ovMZ & = & [ 1 + (\GZ/M_Z)^2 ]^{-\frac{1}{2}} \MZ
&\approx& M_Z&-&34~{\MeV}, 
\label{eqn:mbar}
\\
{\overline \Gamma}_{\Zo} & = & [ 1 + (\GZ / M_Z)^2 ]^{-\frac{1}{2}} \GZ
&\approx& \GZ& - &1~{\MeV}, 
\label{eqn:gbar}
\\
\ovgmu &=& \gmu / (1+i\GZ/MZ). &&
\label{eqn:gmubar}
\end{array}
\renewcommand{\arraystretch}{1.}
\ea
%
\subsection{The OPAL approach}
A completely different approach has been chosen in~\cite{opal}.
With an ad hoc ansatz, the effective couplings of the differential 
cross-section have been
allowed to deviate from what is expected in the Standard
Model in the approximation of effective fermion 
couplings:
\ba
\nonumber
\frac{2s}{\pi\alpha^2} \frac{d \sigma}{d \cos \vartheta} 
&=&
\left(1+\cos\vartheta^2\right)
\left\{
\left|F_A\right|^2 + 8 \Re e \left[F_A^* \chi(s) \right] {\cal C}_{\gamma Z}^s
+ 16 \left|\chi(s)\right|^2 {\cal C}_{ZZ}^s \right\}
\\
& & +~ 2  \cos\vartheta 
\left\{
8 \Re e \left[F_A^* \chi(s) \right] {\cal C}_{\gamma Z}^a
+ 64 \left|\chi(s)\right|^2 {\cal C}_{ZZ}^a \right\},
\ea
with
\ba
\chi(s) &=& \kappa \frac{s}{(s-M_Z^2)^2 + (M_Z \Gamma_Z/s)^2},
\\
\kappa &=& \frac{G_{\mu} M_Z^2}{\sqrt{2} 2 \pi \alpha},
\label{kappagmu}
\ea
and 
\ba
\renewcommand{\arraystretch}{2.0}
\begin{array}{lcl}
{\cal C}_{ZZ}^s &=& 
\displaystyle{
\frac{1}{16} \kappa_{ZZ}^s\left[\sgae+\sgve\right]\left[\sgaf+\sgvf\right]},
\\
{\cal C}_{\gamma Z}^s &=& 
\displaystyle{
\frac{1}{4} \kappa_{\gamma Z}^s \egve \egvf},
\\
{\cal C}_{ZZ}^a &=& 
\displaystyle{
\frac{1}{16} \kappa_{ZZ}^a \egae\egve\egaf\egvf},
\\
{\cal C}_{\gamma Z}^a &=& 
\displaystyle{
\frac{1}{4} \kappa_{\gamma Z}^a \egae \egaf}.
\end{array}
\renewcommand{\arraystretch}{1.8}
\ea
Here, the definitions 
\ba
\label{eqn:effcoup}
\renewcommand{\arraystretch}{1.2}
\begin{array}{lcl}
\egaf & = & \sqrt{\rho}/2, \\
\egvf & = & \egaf(1-4 |Q_f| s_W^{2eff}),
\end{array}
\renewcommand{\arraystretch}{1.y}
\ea 
are used for the effective couplings in the Standard Model.
The parameters $\kappa$ in the definitions of the parameters $\cal C$ allow
for deviations of the cross-sections and asymmetries from the Standard Model
predictions where they are equal to one.
 
The OPAL approach uses the \zfi\ environment with minor modifications.
 
Up to tiny terms which are presumably much smaller than the experimental
accuracy (e.g. the parameter $r_T^{0f}$ in the S-matrix approach),
the OPAL approach is equivalent to the one advocated here.
For the total cross-section and the forward-backward asymmetry the relations
are: 
\ba
\renewcommand{\arraystretch}{2.0}
\begin{array}{lcl}
r_T^f &=& 
\displaystyle{
16 \kappa^2 {\cal C}_{ZZ}^s - 8 \kappa \frac{\Gamma_Z}{M_Z} \Im m F_A
{\cal C}_{\gamma Z}^s},
\\
j_T^f &=& 
\displaystyle{
8 \kappa {\cal C}_{\gamma Z}^s \left[ \Re e F_A +    
\frac{\Gamma_Z}{M_Z} \Im m F_A \right]},
\\
r_{FB}^f &=&  
\displaystyle{
\frac{3}{4} \left\{
64 \kappa^2 {\cal C}_{ZZ}^a - 8 \kappa \frac{\Gamma_Z}{M_Z} \Im m F_A
{\cal C}_{\gamma Z}^a
\right\}},
\\
j_{FB}^f &=& 
\displaystyle{
\frac{3}{4} \left\{
8 \kappa {\cal C}_{\gamma Z}^a \left[ \Re e F_A +    
\frac{\Gamma_Z}{M_Z} \Im m F_A \right]
\right\}}.
\end{array}
\renewcommand{\arraystretch}{1.}
\ea
The additional factors of $\frac{3}{4}$ in the last two relations are due to
the different angular integrations over even
or odd integrands with respect to $\cos \vartheta$.
\subsection {
Effective couplings
\label{oa}
}
Now we define the relation of the model independent approach of \sma\
to the use of effective weak neutral fermion couplings.
The latter is realized in the \zfi\ branch {\tt ZUXSA}.
 
In a simple quantum mechanical interpretation or an approximate quantum field
theoretical one, 
the (complex) residua of the helicity amplitudes \RZfi\ may be 
expressed in terms of effective vector and axial vector couplings of the \Zo\ 
boson to fermions (which are basically real numbers):
\ba
\label{eqn:helva}
\renewcommand{\arraystretch}{1.3}
\begin{array}{lcl}
\RZfn & = & \kappa(\egve+\egae)(\egvf+\egaf), \\
\RZfe & = & \kappa(\egve+\egae)(\egvf-\egaf), \\
\RZfz & = & \kappa(\egve-\egae)(\egvf-\egaf), \\
\RZfd & = & \kappa(\egve-\egae)(\egvf+\egaf), 
\end{array}
\renewcommand{\arraystretch}{1.}
\ea
with the $\kappa$ of~(\ref{kappagmu}) and the couplings \egaf\ and
\egvf\ of~(\ref{eqn:effcoup}).
The parameters $r^f_A$ and $j^f_A$ of~(\ref{eqn:smxs}) are:
\ba
\label{eqn:rcoup}
\renewcommand{\arraystretch}{1.5}
\begin{array}{lclcclcl}
         &   &\rtf   & = & & \displaystyle {\kappa^2 \left[\sgae+\sgve\right]
\left[\sgaf+\sgvf\right]}
 & - & 
2\kappa \egve \egvf C_{Im}, \\
\rlrpolf & = & \displaystyle{\frac{4}{3}} \rfbf & = & & 4\kappa^2\egae
\egve\egaf\egvf
& - & \displaystyle{2 \kappa \egae \egaf C_{Im}}, \\
\displaystyle{\frac{4}{3}} \rfblrf  & = & \rpolf   & = & - & \displaystyle
{2\kappa^2\left[\sgae+\sgve\right]\egaf\egvf} 
& + & \displaystyle{2\kappa \egve \egaf C_{Im}}, \\
\rlrf    & = & \displaystyle{\frac{4}{3}} \rfbpolf & = & - & \displaystyle
{2\kappa^2\egae\egve\left[\sgaf+\sgvf\right]}
  & + & \displaystyle{ 2\kappa \egae \egvf C_{Im}}, \\
\end{array}
\renewcommand{\arraystretch}{1.}
\ea
and
\ba
\renewcommand{\arraystretch}{1.5}
\label{eqn:jcoup}
\begin{array}{lclcclcl}
         &   & \jtf   & = &  & 2 \kappa \egve \egvf (C_{Re} + C_{Im}), & & \\
\jlrpolf & = & \displaystyle{\frac{4}{3}}\jfbf  & = &  & \displaystyle{2 
\kappa \egae \egaf (C_{Re} +
C_{Im})}, & & \\
\displaystyle{\frac{4}{3}}\jfblrf  & = & \jpolf  & = & - & 2 \kappa \egve 
\egaf (C_{Re} + C_{Im}), & & \\
\jlrf    & = & \displaystyle{\frac{4}{3}}\jfbpolf & = & - & \displaystyle
{2 \kappa \egae \egvf 
(C_{Re} + C_{Im})}. & &
\end{array}
\renewcommand{\arraystretch}{1.}
\ea
Additional factors of $(3/4)^{\pm 1}$ are again due to the different
angular integrations for contributions which are even or odd in $\cos
\vartheta$. 

The \gam\Zo-interference is proportional to
$C_{Re}$,
while
$C_{Im}$ are small corrections to it 
and to the resonance peak parameter:
\ba
\renewcommand{\arraystretch}{2.}
\begin{array}{lcl}
C_{Im} & = & \displaystyle{\mid Q_e Q_f\mid \frac{\GZ}{\MZ}\imag
F_A(s)}, \\
C_{Re} & = & \displaystyle{ \mid Q_e Q_f\mid \real F_A(s)}. \\
\end{array}
\renewcommand{\arraystretch}{1.}
\ea
 
\subsection{Model independent QED corrections}
The QED correction factors in \sma\ are universal in the sense	
that they may be used also at energies far away from the \Zo\ peak and in other
approaches. We give here one instructive example for this.
In~\cite{fedo}, the QED corrections to
the total cross-section $\sigma_T(s)$ and the integrated forward-backward
asymmetry have been calculated analytically without 
cuts\footnote 
{The corresponding analytic formulae with a cut on the energy of the 
bremsstrahlung photon may be found in the unpublished Fortran program
ZBIZON.}
to order ${\cal O}(\alpha)$ for reaction~(\ref{e0}).
The \Zo\ exchange cross-section contribution has been presented there
as follows:
\ba
\sigma_T^Z(s)
=
\frac{4\pi\alpha^2}{3s} \left| \chi(s) \right|^2 
\left(v_e^2+a_e^2\right) \left(v_f^2+a_f^2\right) 
\left[1 + \frac{\alpha}{\pi} \left( Q_e^2 \, {\cal H}_0
+Q_e Q_f  \, {\cal H}_4 + Q_f^2 \,  {\cal H}_2 \right) \right].
\label{H}
\ea
In the simplest case (no cuts, neglect of final state masses, no higher
order corrections), the ${\cal H}_2$ e.g. is the well-known
QED final state correction $\frac{3}{4}$.
The ${\cal H}_0$ contains the initial state corrections and ${\cal H}_4$ 
those from the initial final state interference. 
Similar but considerably more involved analytic expressions were derived
for the forward-backward asymmetry.
The following relation holds:
\ba
C_T^r(s) \sim 1 + \frac{\alpha}{\pi} {\cal H}_0(s) + \ldots,
\label{ch}
\ea
where the dots stand for higher order corrections and a potential inclusion
of final state radiation in the $C_T^r$ and the `$\sim$' for potential cuts. 
Following this example, the interested reader may find analogue relations for
the other QED corrections. 
\newpage
\section {Structure of the package}
For the installation of \sma\ the user has
to replace subroutine {\tt BORN} of \zfi\
with subroutine {\tt BORN} of \sma.
 
To run \sma\ one has to initialize first {\zfi} following the
procedure described in~\cite{zfitter}, section~6. 
Then, \sma\ is initialized by a call to subroutine {\tt ASYINIT}. 
Subroutine {\tt ASYTEST} illustrates the initialization procedure and 
performs a comparison of \sma\ with the other model independent 
approaches of \zfi. 
The results are listed in an sample output in appendix~\ref{sec:app}. 

The \sma\ package contains the following interface routines:
\begin{itemize}
\item {\tt SMATASY} -- calculates total cross-sections and asymmetries as 
functions of the center-of-mass energy, the \Zo\ mass and width, the
\Zo- and \gam-exchange terms and the \gam\Zo-interference terms;
\item {\tt SMATRZ} -- calculates  total cross-sections and asymmetries as 
functions of the center-of-mass energy, the \Zo\ mass and width and the
helicity amplitudes;
\item {\tt SMATA01} -- calculates asymmetries as functions of the
center-of-mass energy, the \Zo\ mass and width, the 
\Zo- and \gam-exchange terms and the \gam\Zo-interference terms for
the total cross-sections and the asymmetry parameters, $A^A_0$ and $A^A_1$.
\end{itemize}
Utility routines of interest for the user are:
\begin{itemize}
\item {\tt CORQED} -- calculates the QED correction factors as 
functions of the center-of-mass energy and the \Zo\ mass and width;
\item {\tt RZFRVA} -- calculates the residua of the helicity
amplitudes as functions of
the \Zo\ mass and the effective couplings;
\item {\tt RJFRRZ} -- calculates the \Zo- and \gam-exchange terms
and the \gam\Zo-interference terms as functions of 
the \Zo\ mass and width, the helicity amplitudes and the vacuum polarization;
\item {\tt A01FRRJ} -- calculates the asymmetry parameters, $A^A_0$ and
$A^A_1$, as functions of the \Zo- and \gam-exchange terms and the
\gam\Zo-interference terms;
\item {\tt ASYTRAF} -- performs the transformation of the \Zo\ mass,
width and Fermi's constant between the two definitions in~(\ref{mmggzz}).
\end{itemize}

\begin{table}
\begin{center}
\begin{tabular}{|c|cccccccccccc|} \hline
& & & & & & & & & & & &
\\
final state & \nnbar & \ee & \mm & \tautau & \uubar & \ddbar
& \ccbar & \ssbar & 
\ttbar
& 
\raisebox{0.pt}[2.5ex][0.0ex]{${\bbbar}$}
& $\sum \qqbar$ & Bhabha 
\\ & & & & & & & & & & & &
\\ \hline   
& & & & & & & & & & & &
\\
{\tt INDF}           & 0 & 1 & 2 & 3 & 4 & 5 & 6 & 7 & 8 & 9 & 10 & 11
\\ & & & & & & & & & & & &
\\ \hline
\end{tabular}
\end{center}
\caption{\zfi\ convention for final state labels.
\label{tab:indf}}
\end{table}
 
Although our model independent ansatz implicitly assumes massless fermions 
since it is based on four different helicity amplitudes,  
corrections due to final fermion masses are applied in the sample output in
order to be compatible with \zfi. 
However, the corrections for leptons and light quarks may be
switched off by the \zfi\ flag {\tt POWR}.
The deviations between different branches of \zfi\ itself and of the interface
\sma\ are at most of the order of 
a few tenth of a percent.
The most accurate asymmetry measurement at LEP~1 is expected for the forward
backward asymmetry for leptons at the peak where a theoretical accuracy of 
less than 0.1 \% is demanded. 
The internal deviation between different descriptions 
in the sample output for this quantity is about 0.01 \%. 

\section {Description of the procedures}
 
If not stated differently,  the input and output arguments of the
following subroutines are of the {\tt DOUBLE PRECISION} type.
\subsection {Interface Routines of \sma
\label{sec:int}
}
\subsubsection{Subroutine {\tt SMATASY}
\label{sec:smatasy}
}
Subroutine {\tt SMATASY} is used to calculate total cross-section
and asymmetries as 
functions of $\sqrt{s}$, \ovMZ, \ovGZ, $r^f_A$, $j^f_A$ and $r^{\gamma f}_A$.
The first three coefficients of the Taylor series $r^{f0-2}_A$ 
are also considered.
This refers to the cross-section parameterization~(\ref{eqn:smxs}).
For total cross-sections {\tt SMATASY} is equivalent to the interface
{\tt ZUSMAT}, but note the different definition of \jtf. \\
 
\fbox{\tt CALL SMATASY (INDF,SS,SZMASS,SGAMZ,RT,JT,GT,FT,RA,JA,FA,IA,ASY*)}\\
 
\underline{Input Parameters:}
\begin{itemize}
\item[] {\tt INDF} is the fermion index (see Table \ref{tab:indf})
({\tt INTEGER}).
 
\item[] {\tt SS} is the center-of-mass energy, $\sqrt{s}$, in GeV.
 
\item[] {\tt SZMASS} is the \Zo\ mass, \ovMZ, in GeV.
 
\item[] {\tt SGAMZ} is the total \Zo\ width, \ovGZ, in GeV.
 
\item[] {\tt RT} is the \Zo-exchange term for the total cross-section, \rtf.
 
\item[] {\tt JT} is the \gam\Zo-interference term for the total cross-section,
 \jtf.
 
\item[] {\tt GT} is the \gam-exchange term for the total cross-section,
 $r^{\gam f}_T$.
 
\item[] {\tt FT} is a vector of the first three coefficients $r^{f0-2}_A$,
describing nonresonant contributions.
 
\item[] {\tt RA, JA, FA} are corresponding parameters for a cross-section
difference $\sigma_A^0$.
 
\item[] {\tt IA} defines which cross-section or asymmetry is calculated
({\tt INTEGER})
\footnote{The reserved FORTRAN constants (e.g. {\tt ITOT}) for the
different possible {\tt IA} values are given in the second column below.}:
\ba
\nonumber
\begin{array} {lcllccl}
\mbox{{\tt IA}} & = & \mbox{\tt ITOT}   & = & 0 & \ra & \xstot \\
                &   & \mbox{\tt IFB}    & = & 1 & \ra & \afb  \\
                &   & \mbox{\tt IPOL}   & = & 2 & \ra & \apol  \\
                &   & \mbox{\tt IFBPOL} & = & 3 & \ra & \afbpol \\
                &   & \mbox{\tt ILR}    & = & 4 & \ra & \alr    \\
                &   & \mbox{\tt IFBLR}  & = & 5 & \ra & \afblr  \\
                &   & \mbox{\tt ILRPOL} & = & 6 & \ra & \alrpol 
\end{array}
\ea
\end{itemize}
\underline{Output Parameter:}
\footnote{An asterisk (*) following an argument in a calling sequence
is used to denote an output argument.}
\begin{itemize}
\item[] {\tt ASY} is the total cross-section or an asymmetry selected by
{\tt IA}.
\end{itemize}
 
\subsubsection{Subroutine {\tt SMATRZ}}
\label{sec:smatrz}
 
Subroutine {\tt SMATRZ} is used to calculate total cross-sections
and asymmetries as functions of $\sqrt{s}$, \ovMZ, \ovGZ\
 and the residua of the helicity amplitudes, \RZfi,  as 
introduced in~(\ref{eqn:mat}).
The nonresonant contributions, \Ffin\, in~(\ref{eqn:mat}) are set equal to
zero. According to~(\ref{gamma}), instead of \Rgf\ the vacuum
polarization, $F_A$, is used as free parameter. \\
 
\fbox{\tt CALL SMATRZ (INDF,SS,SZMASS,SGAMZ,RZ0,RZ1,RZ2,RZ3,FA,IA,ASY*)} \\
 
\underline{Input Parameters:}
\begin{itemize}
\item[] {\tt INDF, SS, SZMASS, SGAMZ, IA} have the same meaning as in
        subroutine {\tt SMATASY}, explained in section \ref{sec:smatasy}.
 
\item[] {\tt RZ0, RZ1, RZ2, RZ3} are the residua of the helicity
amplitudes, \RZfi\ ({\tt COMPLEX*16}).
 
\item[] {\tt FA} is the vacuum polarization, $F_A(s)$ ({\tt COMPLEX*16}).
\end{itemize}
 
\underline{Output Parameter:}
\begin{itemize}
\item[] {\tt ASY} has the same meaning as in subroutine {\tt SMATASY}.
\end{itemize}

\subsubsection{Subroutine {\tt SMATA01}}
\label{sec:smata01}
 
Subroutine {\tt SMATA01} is used to calculate asymmetries as  
functions of $\sqrt{s}$, \ovMZ, \ovGZ\, \rtf, \jtf, $r^{\gamma f}_T$
and the asymmetry parameters, $A_0^A$ and $A_1^A$, introduced
in~(\ref{eqn:a0}, \ref{eqn:a1}).
The nonresonant contributions are neglected. \\
 
\fbox{\tt CALL SMATA01 (INDF,SS,SZMASS,SGAMZ,RT,JT,GT,A0,A1,IA,ASY*)} \\
 
\underline{Input Parameters:}
\begin{itemize}
\item[] {\tt INDF, SS, SZMASS, SGAMZ,RT, JT, GT, IA} have the same
meaning as in subroutine {\tt SMATASY}, explained in section \ref{sec:smatasy}.
 
\item[] {\tt A0, A1} are the asymmetry parameters $A_0^A$ and $A_1^A$.
\end{itemize}
 
\underline{Output Parameter:}
\begin{itemize}
\item[] {\tt ASY} has the same meaning as in subroutine {\tt SMATASY}.
\end{itemize}

\subsection {Utility Routines of \sma}
 
\subsubsection{Subroutine {\tt CORQED}}
 
Subroutine {\tt CORQED} calculates $C^\gamma_T$, $C^r_A$, $C^j_A$ and
$C^0_A$ as functions of  $\sqrt{s}$, \ovMZ\ and \ovGZ, according to 
(\ref{eqn:cg}). \\
 
\fbox{\tt CALL CORQED (INDF,SS,SZMASS,SGAMZ,CAR*,CAJ*,CAG*,CA0*,IA)}
 \\
 
\underline{Input Parameters:}
\begin{itemize}
\item[] {\tt INDF}, {\tt SS}, {\tt SZMASS}, {\tt SGAMZ} and {\tt IA}
        have same meaning
        as the parameters explained in Section \ref{sec:int}.
\end{itemize}
 
\underline {Output Parameters:}
\begin{itemize}
\item[] {\tt CAR} is the QED correction factor, $C^r_A$, for the
        \Zo-exchange term, \raf, selected by {\tt IA}.
\item[] {\tt CAJ} is the QED correction factor, $C^j_A$, for the
        \gam\Zo-interference term, \jaf, selected by {\tt IA}.
\item[] {\tt CAG} is the QED correction factor, $C^\gam_A$, for the
        \gam-exchange term, $r^{\gamma f}_A$, selected by {\tt IA}
\footnote{ $C^\gam_A=0$ for $A \neq T$.}.
\item[] {\tt CA0} is the QED correction factor, $C^0_A$, for 
        the nonresonant contribution, $r^{0f}_A$, selected by {\tt IA}.
\end{itemize}

\subsubsection{Subroutine {\tt RZFRVA}}
\label{sec:rzfrva}
 
Subroutine {\tt RZFRVA} is used to calculate
\RZfi\, as functions of \MZ, \gmu, \egve, \egae, \egvf\ and \egaf\ 
using~(\ref{eqn:helva}).
\MZ\ is related to \ovMZ\ by~(\ref{eqn:mbar}).
The subroutine cannot be used for the inclusive hadron channel
({\tt INDF}=10). \\
 
\fbox{\tt CALL RZFRVA (INDF,ZMASS,GMU,GVE,GAE,GVF,GAF,RZ0*,RZ1*,RZ2*,RZ*)} \\
 
\underline{Input Parameters:}
\begin{itemize}
\item[] {\tt INDF} corresponds to the parameter with the same name in
        Section \ref{sec:int}.
 
\item[] {\tt ZMASS} is the mass of the \Zo\ boson, \MZ.
 
\item[] {\tt GMU} is Fermi's Constant, \gmu.
 
\item[] {\tt GVE} is the effective vector coupling of the electron,
        \egve.
 
\item[] {\tt GAE} is the effective axial vector coupling of the electron,
        \egae.
 
\item[] {\tt GVF} is the effective vector coupling of the final-state
        fermion, \egvf.
 
\item[] {\tt GAF} is the effective axial vector coupling of the
        final-state fermion, \egaf.
\end{itemize}
 
\underline{Output Parameters:}
\begin{itemize}
\item[] {\tt RZ0}, {\tt RZ1}, {\tt RZ2}, {\tt RZ3} correspond to
        the parameters explained in section \ref{sec:smatrz}.
\end{itemize}

\subsubsection{Subroutine {\tt RJFRRZ}}
 
Subroutine {\tt RJFRRZ} is used to calculate $r^f_A$, $j^f_A$ and
$r^{\gamma f}_A$ as functions of \ovMZ, \ovGZ, \RZfi\ and
$F_A(s)$, according to~(\ref{eqn:rrjj}).
The subroutine cannot be used for the inclusive hadron channel
({\tt INDF}=10). \\
 
\fbox{\tt CALL RJFRRZ (INDF,SZMASS,SGAMZ,RZ0,RZ1,RZ2,RZ3,FA,RR*,JJ*,GG*,IA)}
 \\
 
\underline{Input Parameters:}
\begin{itemize}
\item[] {\tt INDF}, {\tt SZMASS}, {\tt SGAMZ}, {\tt
        RZ0}, {\tt RZ1}, {\tt RZ2}, {\tt RZ3}, {\tt FA} and {\tt IA}
        have same meaning
        as the parameters explained in Section \ref{sec:int}.
\end{itemize}
 
\underline {Output Parameters:}
\begin{itemize}
\item[] {\tt RR} is the \Zo-exchange term, $r^f_A$, for the cross
        section, $\sigma_A$, selected by {\tt IA}.
\item[] {\tt JJ} is the \gam\Zo-interference term, $j^f_A$, for 
        the cross-section, $\sigma_A$, selected by {\tt IA}.
\item[] {\tt GG} is the \gam-exchange term, $r^{\gamma f}_A$,
        for the cross-section, $\sigma_A$, selected by {\tt IA}.
\end{itemize}

\subsubsection{Subroutine {\tt A01FRRJ}}
 
Subroutine {\tt A01FRRJ} is used to calculate $A_0^A$ and $A_1^A$ 
as functions of \ovMZ, \ovGZ, \rtf, \jtf, $r^{\gamma f}_T$, \raf and
\jaf, according to~(\ref{eqn:a0}, \ref{eqn:a1}). \\
 
\fbox{\tt CALL A01FRRJ (INDF,SZMASS,SGAMZ,RT,JT,GT,RA,JA,A0*,A1*)}
 \\
 
\underline{Input Parameters:}
\begin{itemize}
\item[] {\tt INDF}, {\tt SZMASS}, {\tt SGAMZ}, {\tt
        RT}, {\tt JT}, {\tt GT}, {\tt RA}, and {\tt JA}
        have same meaning
        as the parameters explained in Section \ref{sec:int}.
\end{itemize}
 
\underline {Output Parameters:}
\begin{itemize}
\item[] {\tt A0} and {\tt A1} have the same meaning as the parameters
        explained in Section \ref{sec:int}.
\end{itemize}
 
\subsubsection{Subroutine {\tt ASYTRAF}}
 
Subroutine {\tt ASYTRAF} is used to perform the transformation of
\MZ, \GZ, and \gmu\ from the notations where
the width of the \Zo\ is $\sqrt{s}$ dependent  to the parameters in S-matrix
notation, \ovMZ, \ovGZ, and \ovgmu, according to~(\ref{eqn:mbar}). \\
 
\fbox{\tt CALL ASYTRAF (ZMASS,GAMZ,GMU,SZMASS*,SGAMZ*,SGMU*)} \\
 
\underline{Input Parameters:}
\begin{itemize}
\item[] {\tt ZMASS}  has the same meaning as the parameter in Section
        \ref{sec:rzfrva}.
\item[] {\tt GAMZ} is the total width of the \Zo\ boson, \GZ.
\item[] {\tt GMU} is the Fermi Constant, \gmu.
\end{itemize}
 
\underline{Output Parameters:}
\begin{itemize}
\item[] {\tt SZMASS} and {\tt SGAMZ} have the same meaning as the
       parameters explained in Section \ref{sec:int}.
\item[] {\tt SGMU} is Fermi's Constant, \ovgmu\ ({\tt COMPLEX*16}).
\end{itemize}
 
\subsection*{Acknowledgments}
We would like to thank M.~Gr\"unewald and S.~Riemann
for continuous support and fruitful suggestions and also for a careful reading
of the manuscript 
and D.~Schaile for helpful discussions. 


\newpage
{\Large \bf TEST RUN OUTPUT} 

\label{sec:app}
\vspace {3cm}
\begin{figure}[htb]
\begin{center}
\mbox{\epsfxsize=0.7\textwidth\epsffile {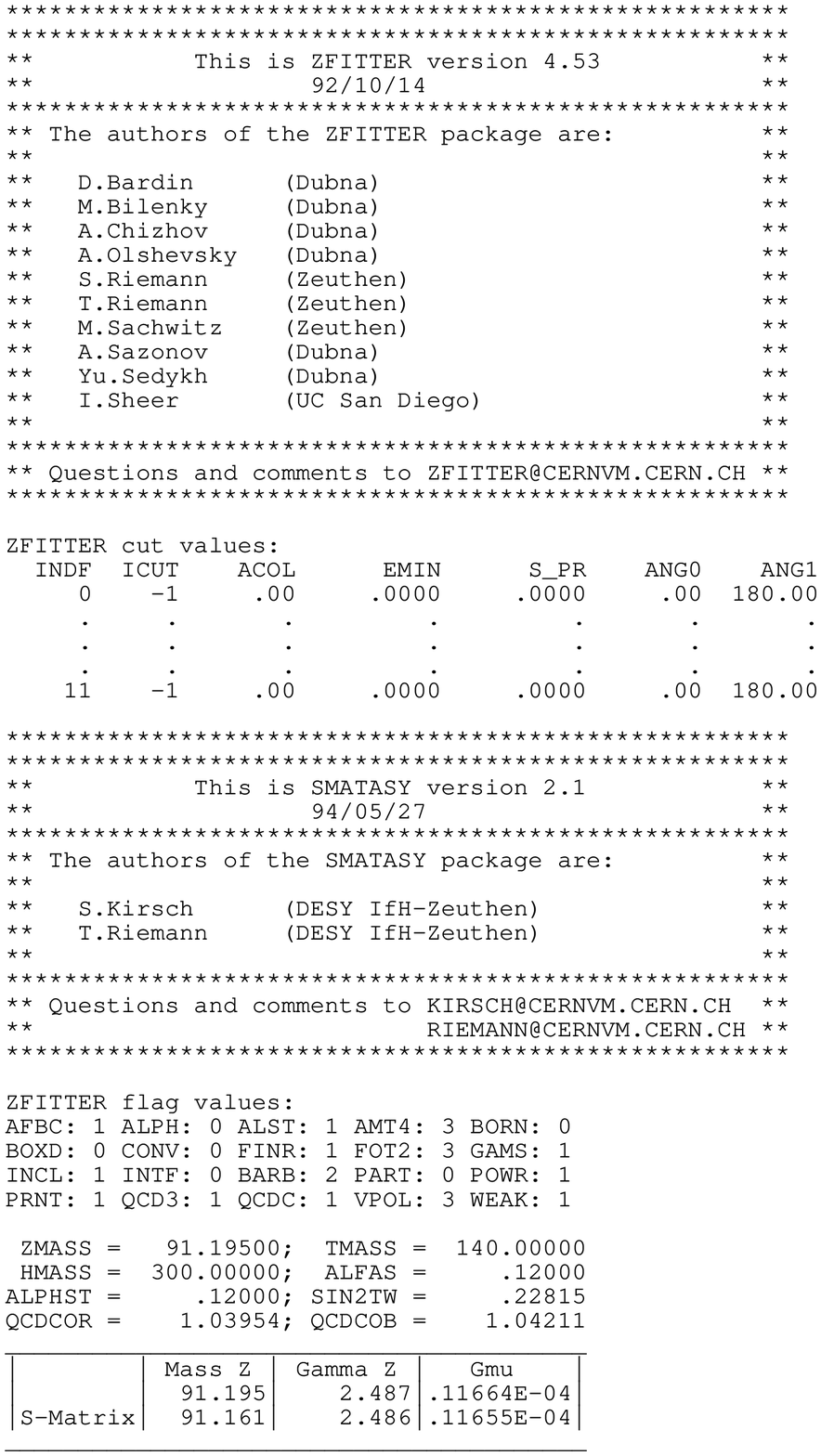}}
\end{center}
\end{figure}
\vskip -2cm

\begin{figure}[p]
\begin{center}
\mbox{\epsfxsize=0.9\textwidth\epsffile {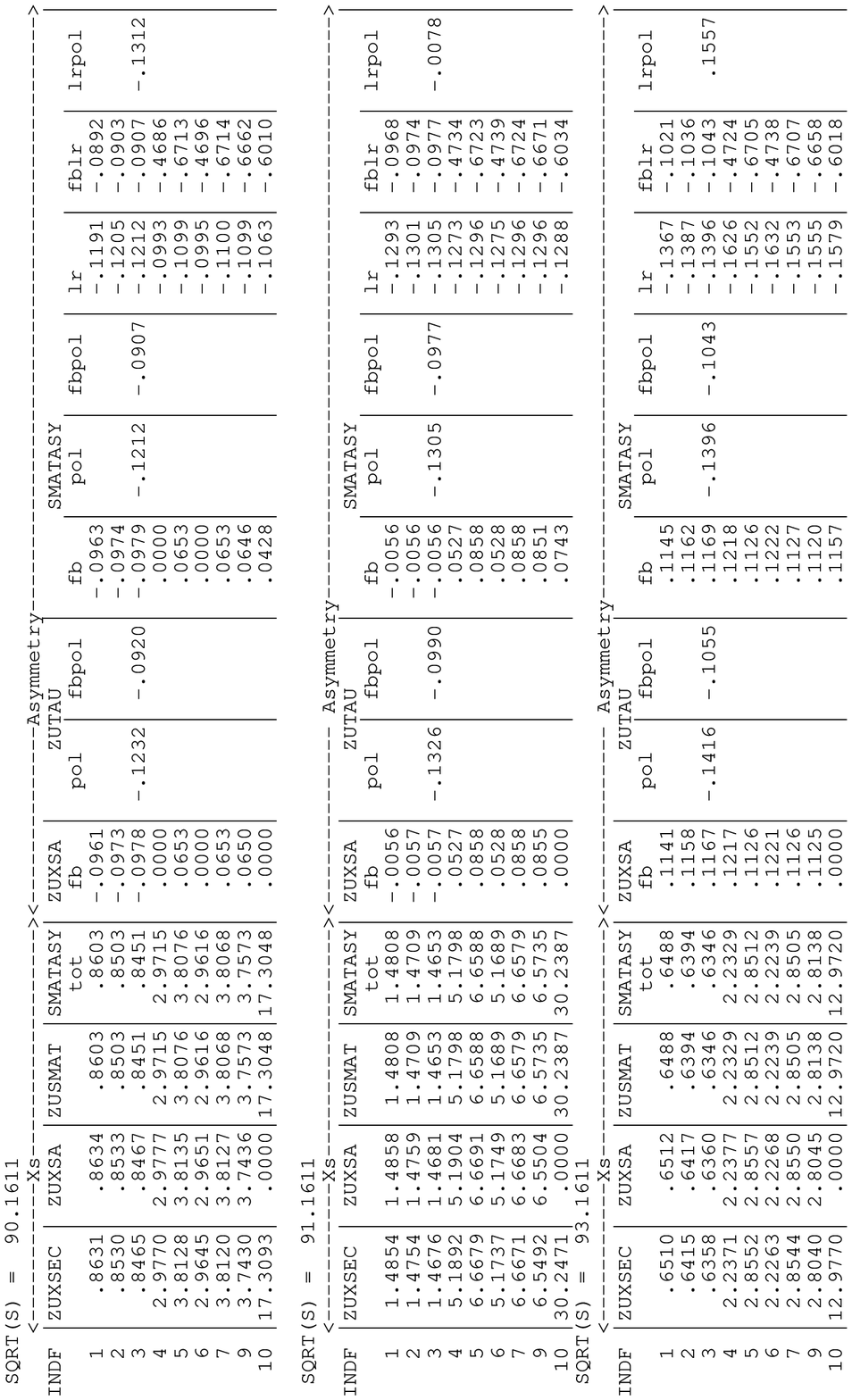}}
\end{center}
\end{figure}

\end{document}